**Numerical simulations of the region of possible sprite inception in the mesosphere above winter thunderstorms under wind shear**

Carynelisa Haspel [a]* and Yoav Yair [b]

[a] The Fredy and Nadine Herrmann Institute of Earth Sciences, Hebrew University of Jerusalem, Jerusalem, Israel

[b] School of Sustainability, Reichman University, Herzliya, Israel

* Corresponding author: Carynelisa Haspel, The Fredy and Nadine Herrmann Institute of Earth Sciences, Hebrew University of Jerusalem, Jerusalem 9190401, Israel;
Email: carynelisa.haspel@mail.huji.ac.il

**Abstract**

Transient luminous events (TLEs) is the collective name given to mesospheric electrical breakdown phenomena occurring in conjunction with strong lightning discharges in tropospheric thunderstorms. They include elves, sprites, haloes and jets, and are characterized by short lived optical emissions, mostly of red (665 nm) and blue (337 nm) wavelengths. Sprites are caused by the brief quasi-electrostatic field induced in the mesosphere, mostly after the removal of the upper positive charge of the thundercloud by a +CG, and they have been recorded above most of the lightning activity centers on Earth. In wintertime, there are just a few areas where lightning occurs, and of those, sprites have been observed over the Sea of Japan, the British Channel, and the Mediterranean Sea. Unlike their summer counterparts, winter thunderstorms tend to have weaker updrafts and as a result, reduced vertical dimensions and compact charge structures, whose positive and negative centers are located at lower altitudes. These storms are often susceptible to significant wind shear and as a result may exhibit a tilted dipole charge structure and a lateral offset of the upper positive charge relative to the main negative charge. We present results of numerical simulations using a three-dimensional explicit formulation of the mesospheric quasi-electrostatic electrical field following a lightning discharge from a typical mid-latitude winter thunderstorm exhibiting tilt due to wind shear and evaluate the regions of possible sprite inception. Our results show, as numerous observations suggest, that sprites can be shifted a large distance from the location of the parent +CG in the direction of the shear and will occur over a larger region compared with non-sheared storms.

**Abbreviations**

December-January-February (DJF)
positive cloud-to-ground strokes (+CG)
transient luminous events (TLEs)
quasi-electrostatic (QE)
charge moment change (CMC)
three dimensional (3D)
impulsive charge moment change (iCMC)
mesoscale convective systems (MCSs)
ELF/VLF (extremely low frequency/very low frequency)
CMCN (charge moment change network)
Severe Thunderstorm Electrification and Precipitation Study (STEPS)

**1. Introduction**

Globally, there are but a few notable locations where winter thunderstorms occur. In northern hemisphere winter (December-January-February; DJF), these span the Sea of Japan, the Gulf stream in the Atlantic, the Indian subcontinent, and the Mediterranean Sea and the neighboring Levant (Christian et al., 2003). Typically, winter thunderstorms have smaller vertical dimensions and lower flash rates, a manifestation of the colder environments in which they develop and the shallower depths of instability. The charge centers are thus located at lower heights and are closer to the surface (Yair et al., 2015). In the presence of wind shear, there may be a lateral displacement of the upper positive charge center from the lower negative one, leading to a tilted-dipole structure (Brook et al., 1982; Levin et al., 1996). This is sometimes manifested by a larger than usual percentage of positive cloud-to-ground strokes (+CG), due to the exposure of the upper charge center to the negative surface below, which facilitates direct discharge from higher altitudes. Still, as was observed by many research groups, these thunderstorms produce very strong lightning (e.g., superbolts; Holzworth et al., 2019) and generate abundant transient luminous events (TLEs; Takahashi et al., 2003; Hayakawa et al., 2004; Adachi et al., 2005; Matsudo et al., 2009; Myokei et al., 2009; Ganot et al., 2007; Yair et al., 2009; Vadislavsky et al., 2009; Pizutti et al., 2022).

The modeling work of Pasko et al. (1995, 1997) identified quasi-electrostatic (QE) fields generated by strong positive cloud-to-ground (+CG) discharge strokes as the mechanism that generates sprites and showed that for low ambient mesospheric conductivities, the decisive factor for the optical intensity of sprites is the value of the removed charge and its altitude, which are expressed by the charge moment change (CMC). Indeed, subsequent works showed that there exists a threshold for CMC below which sprites cannot be produced (Hu et al., 2002;

Asano et al., 2008). For winter sprites, Adachi et al. (2004) found that the number of column sprite elements is related to the peak current, while the length of each sprite element is proportional to the CMC of the parent flash.

Despite the increasing number of observations of TLEs by ground- and space-based platforms, some fundamental issues still remain poorly understood, among which are the possible temporal delay between the sprite and the causative stroke and the possible lateral offset of the sprite from the location of the causative stroke, the latter of which can reach tens of kilometers (São-Sabbas et al. 2003). Here we present results of numerical modeling of the quasi-electrostatic electric field in the mesosphere following a +CG discharge in winter thunderstorms developing under wind shear and show that the region of possible sprite inception is displaced under such conditions, depending on the value of the shear and the properties of the discharge.

## 2. Methods
### 2.1 The model

As mentioned in section 1, one of the unique factors that we investigate here is the effect of wind shear on the region of possible sprite inception above a thunderstorm. Such wind shear creates a lack of lateral symmetry in the cloud charge configuration. Likewise, consecutive discharges from neighboring cloud cells can create a lack of lateral symmetry in the charge configuration. Thus, for the simulations in the present study, we adopt the three dimensional (3D) quasi-electrostatic model presented by Haspel et al. (2022), which was developed specifically to handle non-symmetric charge configurations in a large 3D domain.

A number of previous theoretical studies of sprite development have employed models whose predominant numerical scheme is based on a finite differencing. Due to the combined requirement of a large simulation domain and sufficient time resolution to properly capture the changes in the electric field, the simulations conducted with such models are often two-dimensional axisymmetric simulations (e.g., Pasko et al., 1995, 1997, 1999; Pulinets et al., 2000; Thomas et al., 2005; Hu and Cummer, 2006; Hu et al., 2007; Riousset et al. 2007; Kreibel et al., 2008; Riousset et al. 2010; Mallios and Pasko, 2012; Liu et al., 2015; Malagón-Romero et al., 2019). See the explanation in Riousset et al. (2010). In contrast, the model of Haspel et al. (2022) is based on an analytical solution of Poisson's equation using the method of images with respect to both an upper and a lower boundary. As such, the model of Haspel et al. (2022) is able to handle non-symmetric charge configurations in three dimensions over a large domain and sufficient time resolution with the same efficiency as it handles symmetric charge configurations. Moreover, for the same 3D charge configuration, the model of Haspel et al. (2022) is more numerically stable, more accurate, and less sensitive to the choice of spatial resolution than finite-differencing schemes. In addition, the model of Haspel et al. (2022) does not necessitate any artificial side boundary conditions, and it is readily parallelizable on multiple processors.

See Haspel et al. (2022) for the full details of the model, but the basic structure of the model is as follows. A 3D Cartesian coordinate system is used, and the computational domain extends vertically from the Earth's surface (altitude $z = 0$), presumed to be perfectly conducting with zero electric potential, to the base of the ionosphere ($z_{\text{ionosphere}}$), also presumed to be perfectly conducting with zero electric potential. A value of $z_{\text{ionosphere}} = 90$ km is used as the default, but the sensitivity of the results to the value of $z_{\text{ionosphere}}$ is also examined. The total charge in Coulombs for charge $i$ positioned at the point $(x_i, y_i, z_i)$ at time $t$ is symbolized $Q_{\text{tot}_i}(x_i, y_i, z_i, t)$, and this total charge comprises any cloud charge centers, $Q_{\text{cloud}}$, that exist at that point plus any free charges, $Q_{\text{free}}$, that exist at that point. (See below.)

The components of the electric field at all points in the domain, $\vec{E}(x, y, z, t)$, are calculated analytically, as follows:

$$E_x(x,y,z,t) = \frac{1}{4\pi\varepsilon_0} \sum_{i=1}^{N_{\text{charges}}} \sum_{n=-n_{\max}}^{n_{\max}} \left[ \frac{Q_{\text{tot}_i}(x_i,y_i,z_i,t)\cdot(x-x_i)}{\left((x-x_i)^2+(y-y_i)^2+\left(z-(2nz_{\text{upper boundary}}+z_i)\right)^2\right)^{3/2}} \right.$$

$$\left. - \frac{Q_{\text{tot}_i}(x_i,y_i,z_i,t)\cdot(x-x_i)}{\left((x-x_i)^2+(y-y_i)^2+\left(z-(2nz_{\text{upper boundary}}-z_i)\right)^2\right)^{3/2}} \right]$$

$$E_y(x,y,z,t) = \frac{1}{4\pi\varepsilon_0} \sum_{i=1}^{N_{\text{charges}}} \sum_{n=-n_{\max}}^{n_{\max}} \left[ \frac{Q_{\text{tot}_i}(x_i,y_i,z_i,t)\cdot(y-y_i)}{\left((x-x_i)^2+(y-y_i)^2+\left(z-(2nz_{\text{upper boundary}}+z_i)\right)^2\right)^{3/2}} \right.$$

$$\left. - \frac{Q_{\text{tot}_i}(x_i,y_i,z_i,t)\cdot(y-y_i)}{\left((x-x_i)^2+(y-y_i)^2+\left(z-(2nz_{\text{upper boundary}}-z_i)\right)^2\right)^{3/2}} \right]$$

$$E_z(x,y,z,t) = \frac{1}{4\pi\varepsilon_0} \sum_{i=1}^{N_{\text{charges}}} \sum_{n=-n_{\max}}^{n_{\max}} \left[ \frac{Q_{\text{tot}_i}(x_i, y_i, z_i, t) \cdot (z - (2nz_{\text{upper boundary}} + z_i))}{\left((x-x_i)^2 + (y-y_i)^2 + (z - (2nz_{\text{upper boundary}} + z_i))^2\right)^{3/2}} \right.$$

$$\left. - \frac{Q_{\text{tot}_i}(x_i, y_i, z_i, t) \cdot (z - (2nz_{\text{upper boundary}} - z_i))}{\left((x-x_i)^2 + (y-y_i)^2 + (z - (2nz_{\text{upper boundary}} - z_i))^2\right)^{3/2}} \right]$$

, (1)

where $\varepsilon_0$ is the permittivity of free space, $N_{\text{charges}}$ is the total number of points containing charges in the domain, and $n$ is an integer ranging from $-n_{\max}$ to $n_{\max}$. As explained in Haspel et al. (2022), a numerical procedure presented by Binney and Tremaine (1987) for solutions to Poisson's equation for gravitational potential is adapted to compute the $n = 0$ term in Equation 1 for points that coincide with a nonzero value of charge, i.e., where $(x, y, z) \rightarrow (x_i, y_i, z_i)$.

The evolution of the cloud charge centers in space and time is prescribed (see section 2.2), while the evolution of the free charges in space and time is dictated by the equation of charge continuity/charge conservation (see, e.g., Pasko et al. (1995) and Riousset et al. (2010)):

$$\frac{\partial \rho_{\text{free}}(x,y,z,t)}{\partial t} + (\nabla \sigma(x,y,z)) \cdot \vec{E}(x,y,z,t) + \frac{\sigma(x,y,z) \rho_{\text{tot}}(x,y,z,t)}{\varepsilon_0} = 0, \qquad (2)$$

where $\rho_{\text{free}}(x, y, z, t)$ is the density of free charges, $\rho_{\text{tot}}(x, y, z, t)$ is the total charge density, which we obtain by dividing $Q_{\text{tot}}(x, y, z, t)$ by the volume of a grid box ($\Delta x \cdot \Delta y \cdot \Delta z$), and $\sigma(x, y, z)$ is the ambient electric conductivity. Equation 2 is discretized with a forward-Euler-style method with respect to time:

$$\rho_{\text{free}}(x,y,z,t+\Delta t) = \rho_{\text{free}}(x,y,z,t) - \left[ (\nabla \sigma(x,y,z)) \cdot \vec{E}(x,y,z,t) - \frac{\sigma(x,y,z) \rho_{\text{tot}}(x,y,z,t)}{\varepsilon_0} \right] \cdot \Delta t$$

. (3)

A higher-order, Adams-Bashforth-style method of time discretization (see, e.g., Butchner, 2003) was also tested, but the results were no more robust than the simple Euler-style method of Equation 3. Thus, we use Equation 3 throughout this study.

In each time step, Equation 1 is used to calculate $\vec{E}(x,y,z,t)$, then Equation 3 is used to update the values of $\rho_{\text{free}}(x,y,z,t)$ based on the values of $\vec{E}(x,y,z,t)$ from Equation 1, then the new values of $\rho_{\text{free}}(x,y,z,t)$ are multiplied by $\Delta x \cdot \Delta y \cdot \Delta z$ to produce new values of $Q_{\text{free}}(x,y,z,t)$, and then the new values of $Q_{\text{free}}(x,y,z,t)$ are added to the values of $Q_{\text{cloud}}(x,y,z,t)$ appropriate for the same time step (see section 2.2) to produce new values of $Q_{\text{tot}_i}(x_i,y_i,z_i,t)$. Finally, the values of $\vec{E}(x,y,z,t)$ are zeroed out, and Equation 1 is used once again to calculate new values of $\vec{E}(x,y,z,t)$ from the new values of $Q_{\text{tot}_i}(x_i,y_i,z_i,t)$.

For simplicity, as in Haspel et al. (2022), in the present study, we choose $\sigma(x,y,z)$ to equal the ion conductivity represented by profile A of Pasko et al. (1997), i.e., $\sigma(x,y,z) = 5.0 \times 10^{-14} \cdot \exp(z/6 \text{ km})$ S m$^{-1}$, independent of time. This profile is similar to the ambient conductivity profiles at time $t = 0.501$ s of the simulations shown in Figure 3 of Pasko et al. (1995) for cloud charge center values of 100-200 C and of the simulations shown in Figure 12 of Pasko et al. (1997) for cloud charge center values of 50-150 C, as well as the ambient conductivity profile shown in Figure 1c of Hiraki and Fukunishi (2006). However, we also discuss the sensitivity of our results to the ambient conductivity profile.

As demonstrated in Haspel et al. (2022), a grid spacing of $\Delta x = \Delta y = 4$ km and $\Delta z = 1$ km, a value of $n_{\max} = 1$ in Equation 1, and a time step of $\Delta t = 0.0001$ s in Equation 3 are sufficient for the simulations with the simulation parameters used here. Likewise, as demonstrated in Haspel et al. (2022), a threshold value of charge used in the summation over $i$ in Equation 1 of $Q_{\text{threshold}} = 0.000008$ C is sufficient for the simulations with the simulation parameters used here. We use these settings by default.

In addition, by default, we set the horizontal bounds of the domain to be $x_{\min} = -60$ km, $x_{\max} = 60$ km, $y_{\min} = -60$ km, and $y_{\max} = 60$ km, but we increase horizontal bounds as necessary for simulations with wider charge distributions.

We define the region(s) of possible sprite inception at a given time $t$ to consist of any points in the mesospheric region of the domain at which the magnitude of $\vec{E}(x, y, z, t)$ exceeds the conventional electrical breakdown field, $E_k$, at that point. For our winter thunderstorm simulations, by default, we prescribe $E_k$ as $E_k(z) = 3.2 \times 10^6 \cdot N(z)/N(z=0)$ V m$^{-2}$, with $N(z)$ corresponding to average January conditions at a latitude of 30°N, using the extension of the US Standard Atmosphere, 1976 (Minzner, 1977) presented in Champion et al. (1985). For contrasting the winter storm with a summer storm, for the summer storm, we use $N(z)$ corresponding to average July conditions at a latitude of 30°N. (See Figure 1 in Haspel et al. (2020) for plots of these profiles.)

## 2.2 The thunderstorm electrical configuration

For our default eastern Mediterranean winter storm cloud, we place the positive cloud charge center at $z_{\text{upper}} = 9$ km altitude and the negative cloud charge center at $z_{\text{lower}} = 6$ km altitude, but we test additional values of cloud charge center altitude as well. This choice of default altitudes reflects the average heights of the freezing level and the −20°C and −40°C isotherms in typical winter thunderstorms in Israel, based on radiosonde data obtained from the Israeli Meteorological Service in Bet-Dagan. We examined the heights for selected days when thunderstorms were present over the Mediterranean Sea off the coast of Israel and sprites were optically detected (Vadislavsky et al. 2009). As in Pasko et al. (1995, 1997) and in subsequent studies adopting a similar modeling framework, and as in Haspel et al. (2022), the positive and negative cloud charge centers, respectively, are each spread spatially according to a time-dependent Gaussian function:

$$Q_{\text{cloud}_i}(x,y,z,t) = Q_0(t)\exp\left[-\frac{(x-x_i)^2+(y-y_i)^2+(z-z_i)^2}{a^2}\right]. \quad (4)$$

For our default eastern Mediterranean winter storm cloud, we set the radius of the Gaussian to be $a = 1$ km, rather than the 3-km radius used in Pasko et al. (1995, 1997) and Haspel et al. (2022). This choice of $a = 1$ km is based on the fact that cumulonimbus clouds in winter thunderstorms tend to exhibit more compact lateral and vertical dimensions as compared with their summer counterparts (see, e.g., Figure 8 of Yair et al. (2015)). As in Haspel et al. (2022), Equation 4 is normalized such that the sum of the values of the exponential function over the Gaussian "disk" of the upper and lower cloud charge centers, respectively, is equal to 1.

As in Pasko et al. (1995, 1997) and Haspel et al. (2022), $Q_0(t)$ is prescribed in three stages. In stage 1, the positive and negative cloud charge centers build up due to microphysical processes in the cloud (Yair, 2008) with a prescribed time dependence. In stage 2, by default, a positive cloud-to-ground lightning discharge (+CG) occurs, and the positive charge center is removed from the cloud with a prescribed time dependence. In stage 3, only the negative cloud charge center remains in the cloud for the remainder of the simulation.

In contrast to Pasko et al. (1995, 1997) and Haspel et al. (2022), the total discharge time of a lightning event in the present study is taken to be 10 ms rather than 1 ms. We base this choice on the fact that sprite-producing +CG strokes often exhibit long continuing currents (Cummer et al., 2005), where values of up to 150 ms have been reported. Additionally, Lyons et al. (2008) noted that a CMC of 500 C km is a reliable metric for sprite generation (in 50% of the cases), and it is achieved after 10 ms of return stroke current, with a variability in the range of 1-100 ms. Thus, a 10-ms discharge time seems to be a conservative lower limit for sprite-producing +CG strokes. To achieve a total discharge time of $t_{\text{total discharge}} = 10$ ms, for consistency, we begin the discharge with the same tanh dependence as that in Pasko et al. (1995, 1997) and Haspel et al. (2022) with a timescale of $\tau_{\text{tanh}} = 0.001$ s, but then after 0.0005 s, we decrease the pace of the discharge to linear in time for the remainder of the discharge. For the positive charge center:

$$Q_0(t) = +Q_{0_{\max}} \cdot \frac{\tanh\left(\frac{t}{0.5000 \text{ s}}\right)}{\tanh(1)}, \quad 0.0000 \text{ s} \leq t < 0.5000 \text{ s}$$

$$Q_0(t) = +Q_{0_{\max}} \cdot \left[1 - \frac{\tanh\left(\frac{t - 0.5000 \text{ s}}{\tau_{\text{tanh}}}\right)}{\tanh(1)}\right], \quad 0.5000 \text{ s} \leq t \leq 0.5005 \text{ s} \quad (5)$$

$$Q_0(t) = Q_0(t = 0.5005 \text{ s}) - \left[\frac{Q_0(t = 0.5005 \text{ s})}{0.0095 \text{ s}}\right] \cdot [t - 0.5005 \text{ s}], \quad 0.5005 \text{ s} < t \leq 0.5100 \text{ s}$$

$$Q_0(t) = 0 \text{ C}, \quad 0.5100 \text{ s} < t \leq 1.0000 \text{ s}$$

and for the negative charge center:

$$Q_0(t) = -Q_{0_{\max}} \cdot \frac{\tanh\left(\frac{t}{0.5000 \text{ s}}\right)}{\tanh(1)}, \quad 0.0000 \text{ s} \leq t < 0.5000 \text{ s}, \quad (6)$$

$$Q_0(t) = -Q_{0_{\max}}, \quad 0.5000 \text{ s} \leq t \leq 1.0000 \text{ s}$$

where $Q_{0_{\max}}$ is the maximum absolute value of both the positive charge center and the negative charge center. See the dotted magenta curves and the blue curves, respectively, in Figure 1. Our default value of $Q_{0_{\max}}$ is 200 C, but we show the sensitivity of the results to variations in the value of $Q_{0_{\max}}$, and for our simulations of consecutive discharges in neighboring cloud cells, we set $Q_{0_{\max}}$ to 100 C in each cloud.

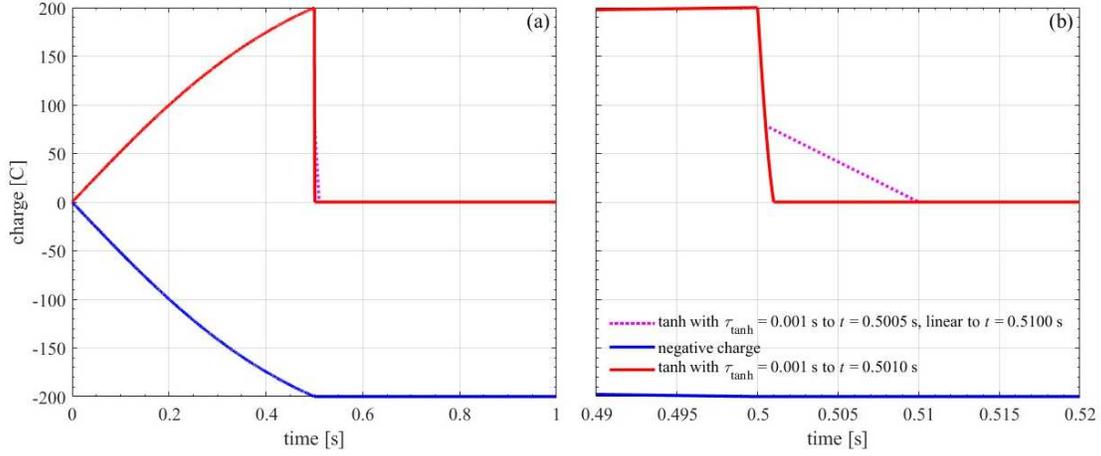

**Figure 1.** Time dependence of the upper positive and lower negative cloud charge centers for the simulations in this study. The default time dependence of the upper positive charge center is given by the dotted magenta curves, and the time dependence of the lower negative charge is given by the solid blue curves. (a) Full time dependence; (b) a zoom in on the time dependence near and during the discharge.

### 2.3 Values of shear

In order to assess the wind shear in winter thunderstorms producing sprites, we inspected nights in the 2007-2008 winter season reported by Vadislavsky et al. (2009). Bet-Dagan radiosonde hodograph data for 00UT (0200 local time) were used to compute the difference in wind speed between 6 and 9 km, which are the average altitudes of the −20°C and −40°C isotherms, where the negative and positive charge centers reside (Krehbiel et al., 1986; Yair et al., 2015). The wind shear was computed by subtracting the lower altitude wind speed from the upper altitude wind speed and dividing by the depth of the layer between the two isotherms. The lowest shear value found was 4.15 m s$^{-1}$ km$^{-1}$, and the highest shear value found was 16.67 m s$^{-1}$ km$^{-1}$, with values in the range of 5-6 m s$^{-1}$ km$^{-1}$ being the most frequent ones for the sampled events.

We label the offset of the positive upper cloud charge center in the eastward direction (i.e., the offset in the positive $x$-direction) with the term "tilt", and we obtain the values of tilt using the following formula:

$$\text{tilt [km]} = \text{shear [m s}^{-1}\text{ km}^{-1}\text{]} \times \left(z_{\text{upper}} - z_{\text{lower}}\right) \text{ [km]} \times \Delta t_{\text{mature charge structure}} \text{ [s]} \times \frac{1}{1000} \text{ [km m}^{-1}\text{]}$$

, (7)

where $\Delta t_{\text{mature charge structure}} = 300$ s is the time span for mature charge structure obtained by assuming a typical winter cloud updraft speed of 10 m s$^{-1}$. Note that this time span need not match the timescale of 0.5 s for the "numerical" buildup of cloud charge and screening charge used in Equations 5-6 above. Based on the above, the values of wind shear (rounded to the nearest 0.01 m s$^{-1}$ km$^{-1}$) and the corresponding values of tilt (rounded to the nearest 0.1 km) used in this study are listed in Table 1.

**2.4 Values of impulse charge moment change (iCMC)**

As part of an operational procedure prepared for the SPRITES 2007 campaign and following the work of Cummer and Lyons (2005), a reliable parameter was sought that could be determined in real time and could aid in targeting sprite-producing cells within the large mesoscale convective systems (MCSs) in the observed area. That parameter was determined to be the impulse charge moment change (iCMC), which was based on the ELF/VLF (extremely low frequency/very low frequency) radiation emitted by the discharge process in the first 2 ms of the return stroke. It proved to be an accurate predictor for sprite-producing +CG strokes, with greater than 50% success and a low false alarm rate (Lyons et al., 2008). An iCMC value of ~300 C km was determined to be a high probability indicator for the emergence of sprites. It should be noted that Cummer et al. (2013) analyzed over 14 million lightning strokes detected by the CMCN (charge moment change network) and found only a weak correlation between the stroke peak current and the value of the iCMC.

The value of iCMC for a lightning discharge in a given cloud cell in our simulations is calculated according to its definition as $\text{iCMC} = \left[ Q_{0_{max}} - Q_0(t = 0.002 \text{ s}) \right]$ [C] $\cdot z_{upper}$ [km] and is rounded to the nearest whole number in units of C km.

**3. Results**

**3.1 Vertical profiles of charge and electric field for our default winter thunderstorm**

The vertical profiles and vertical cross sections of charge density and electric field in center of the domain for our default winter thunderstorm are shown in Figure 2. (See Table 2, simulation 1 for a summary of the parameters.) Similar to the results shown in Figures 1c and 2 of Pasko et al. (1995), in Figure 11 of Pasko et al. (1997), and in Figure 1 of Haspel et al. (2022), from Figure 2, we see a clear increase in the amplitude of the electric field in the mesosphere at $t = 0.501$ s and a correspondingly clear increase in the free charge density in the mesosphere at that same time step (red curves in the row 1 of the plots; shaded contours in row 3 of the plots). Note that the absolute peak of the magnitude of the electric field and the free charge density does not actually occur at $t = 0.501$ s in the present simulation.

Nevertheless, we display this time step in Figure 2 for direct comparison with Pasko et al. (1995, 1997) and Haspel et al. (2022). Likewise, as can be seen in rows 2-4 of the plots of Figure 2, there is a clear lack of symmetry in the charge density and in the magnitude of the electric field, and this is due to the wind shear. This lack of symmetry results in peak values of charge density and electric field that are not in the center of the domain. Nevertheless, in the row 1 of the plots, we display the vertical profiles of charge density and electric field in the center of the domain for direct comparison/contrast with Pasko et al. (1995, 1997) and Haspel et al. (2022). The lack of symmetry in the charge density and in the magnitude of the electric field is exactly the type of phenomenon that can be captured only with a fully 3D simulation such as ours rather than with a 2D axisymmetric model.

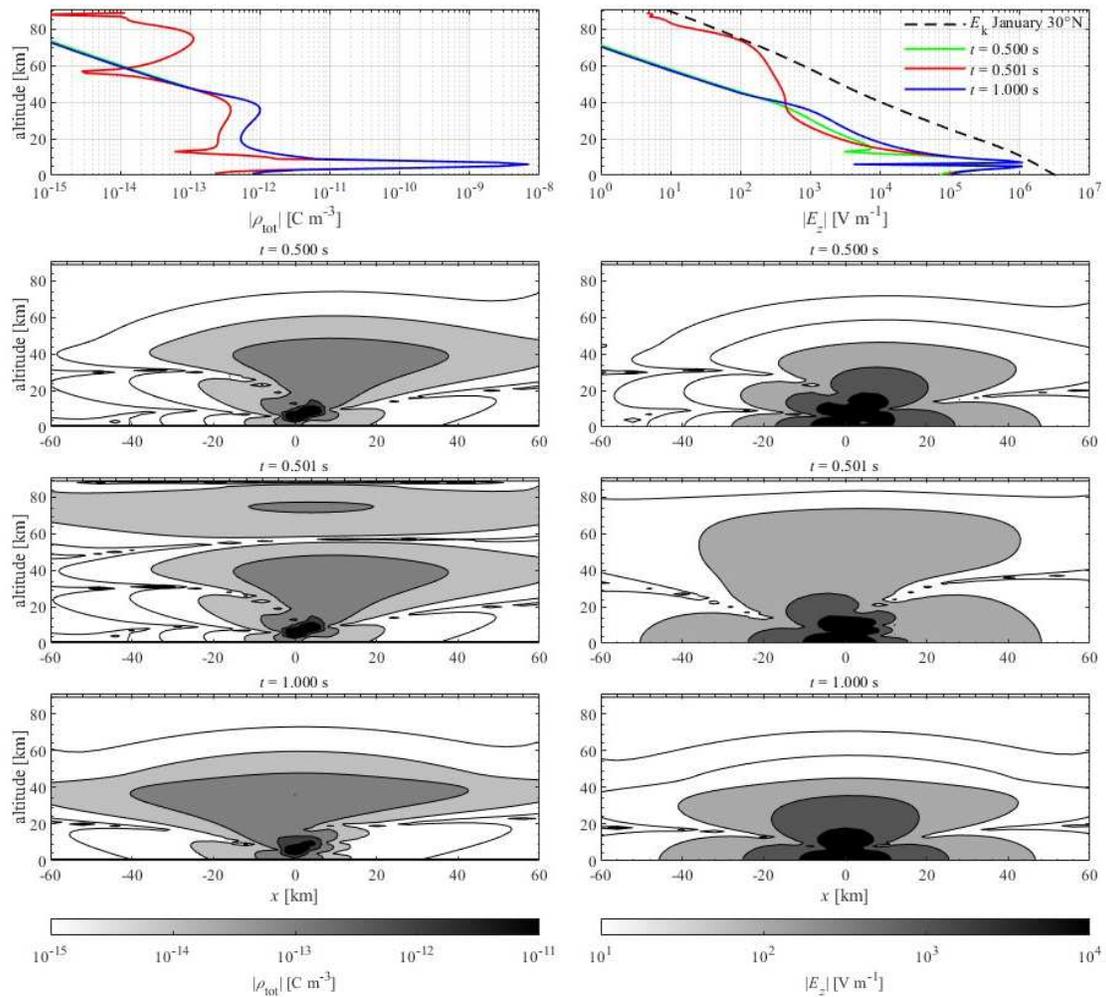

**Figure 2.** Simulation of the vertical profile of total charge density and the $z$-component of the electric field in a cross section of the domain at $y = 0$ km for our default winter thunderstorm with the parameter set given by Table 2, simulation 1.

## 3.2 Winter versus summer

As explained in section 2.1, we define the region(s) of possible sprite inception at a given time $t$ to consist of any points in the mesospheric region of the domain at which the magnitude of $\vec{E}(x,y,z,t)$ exceeds the conventional electrical breakdown field, $E_k$, at that altitude. To illustrate where these regions exist, we plot contours of the difference between $|\vec{E}(x,y,z,t)|$ and $E_k(z)$ in a cross section of the domain at $y=0$ km; the regions of possible sprite inception are then the zero contours. The results for our default winter thunderstorm (refer to Table 2, simulation 1) are given by the blue contours in Figure 3. For comparison/contrast, the results for a summer thunderstorm with the parameters used in Figure 1 of Haspel et al. (2022) but this time with the $E_k$ profile of July (see Table 2, simulation 2), are given by the red contours in Figure 3. Note that there are a total of six differences between the winter and summer parameter sets used for the simulations displayed in Figure 3: (1) the altitudes of the positive and negative cloud charge centers, (2) the radii of the cloud charge centers, (3) the tilt or lack of tilt, (4) the total discharge time, (5) the shape of the discharge time dependence, and (6) the vertical profile of $E_k$.

From Figure 3, we can see that at times $t=0.5001$-$0.5005$ s, the region of possible sprite inception for the winter storm is shifted by ~4.5 km but is not much different in size from the region of possible sprite inception for the summer storm despite the lower iCMC. (The winter $E_k$ profile as opposed to the summer $E_k$ profile partially compensates for the lower iCMC.) In addition, at times $t=0.5003$-$0.5005$ s, the region of possible sprite inception for the winter storm extends down to slightly lower altitudes; this might also be due to the differing $E_k$ profile. On the other hand, from $t=0.5006$ s, the region of possible sprite inception for the winter storm is smaller than that of the summer storm and lasts only until $t=0.5008$ s, i.e., lasts only 0.0008 s after the onset of the discharge. This is mostly due to the fact that the rapid decrease in charge during the discharge lasts only 0.0005 s in the case of the default winter storm in contrast to lasting 0.001 s in the case of the summer storm. The fact that the region of possible sprite inception for the winter storm is smaller and exists for fewer time steps than that of the summer storm supports the fact that there are fewer sprites observed over winter thunderstorms than over summer thunderstorms. (Williams and Yair, 2006; Evtushenko et al., 2022).

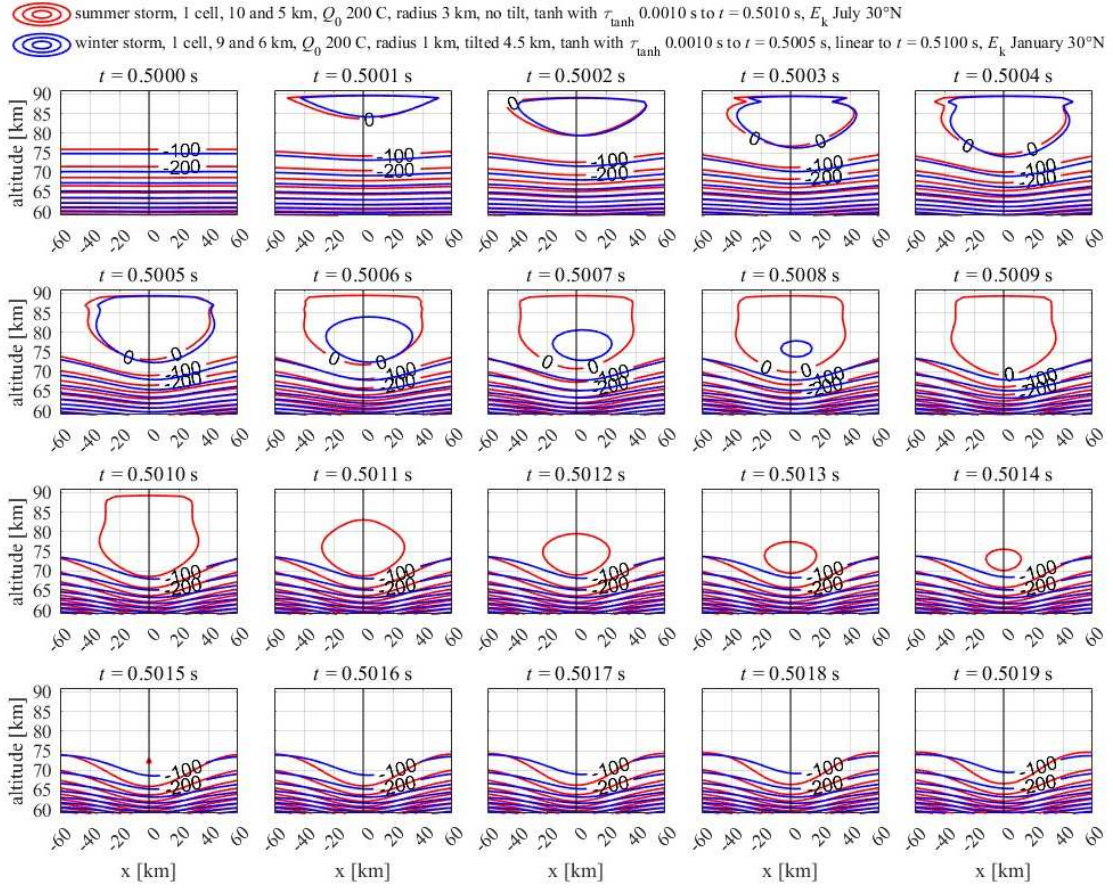

**Figure 3.** Contours of the difference between $|\vec{E}(x,y,z,t)|$ and $E_k(z)$ in a cross section of the domain at $y = 0$ km for a summer thunderstorm with the parameters used in Figure 1 of Haspel et al. (2022) but this time with $E_k$ profile of July (red contours; Table 2, simulation 2) and for our default winter thunderstorm (blue contours; Table 2, simulation 1).

### 3.3 Minimum charge that produces a region of possible sprite inception

The value of iCMC for a given amount of accumulated charge should be smaller in winter thunderstorms due to the lower altitudes of the charge centers, and by inference, the lower discharge heights of sprite-producing +CG strokes. Indeed, Hu et al. (2002) showed a minimum CMC value of 120 C km for sprite inception, but that value was based on ELF data obtained during the Severe Thunderstorm Electrification and Precipitation Study (STEPS) campaign in the summer of 2000, and so does not represent wintertime conditions. Given that, here we perform a sensitivity test, seeking the minimum charge that produces a region of possible sprite inception within our winter storm parameters. For these simulations, we hold all other parameters constant and vary $Q_{0_{max}}$ from 200 C (the default value) down to 100 C.

The values of $Q_{0_{max}}$ and the corresponding values of iCMC, along with all of the other parameters, are listed in Table 2, simulation set 3. The results are shown in Figure 4. We can see that as the value of $Q_{0_{max}}$ decreases and the value of iCMC decreases correspondingly, the regions of possible sprite inception (again, the zero contours) contract. When $Q_{0_{max}}$ = 100 C, there is only a tiny region of possible sprite inception at $t$ = 0.5001 s (the black zero contour). Thus, we find that $Q_{0_{max}}$ = ~100 C is the minimum charge that produces a region of possible sprite inception within our winter storm parameters, with a respective iCMC value of 602 C km.

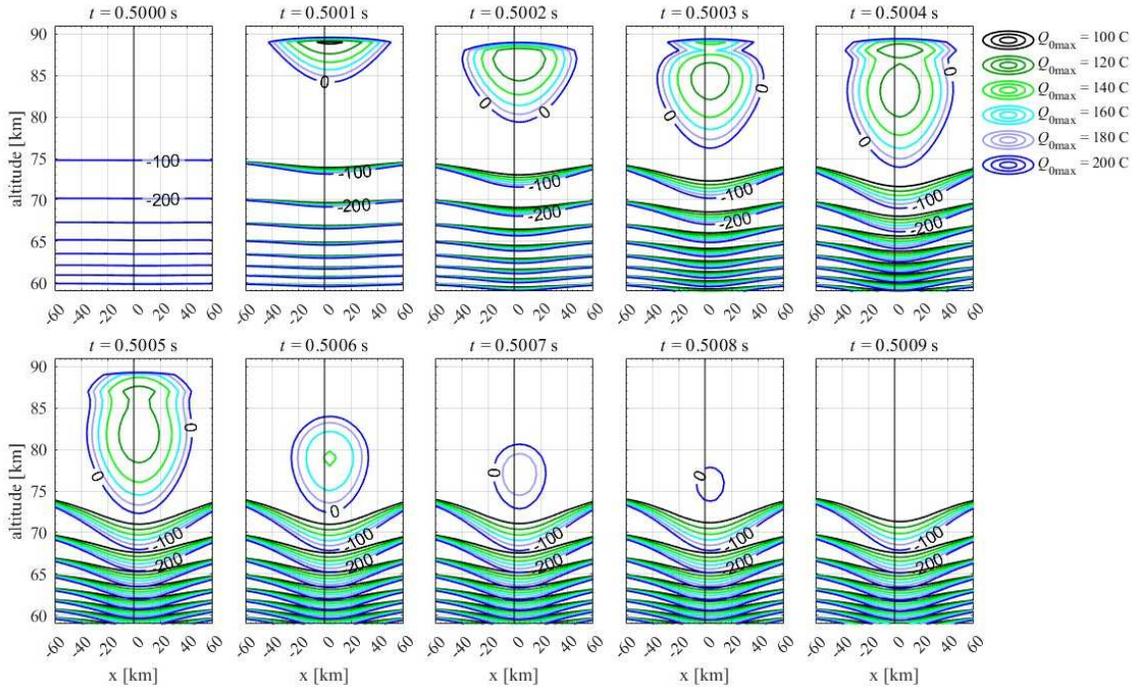

**Figure 4.** Contours of the difference between $\left|\bar{E}(x,y,z,t)\right|$ and $E_k(z)$ in a cross section of the domain at $y$ = 0 km for a winter thunderstorm with varying values of $Q_{0_{max}}$ and with all other parameters held constant at their default values (Table 2, simulation set 3).

### 3.4 The effect of wind shear and sensitivity to the values of wind shear

As described in section 2.3, various values of wind shear and therefore cloud tilt can be extracted from radiosonde data of eastern Mediterranean winter storms, with values of tilt ranging from 3.7 km to 15.0 km (refer to Table 1). Here we examine the sensitivity of the location and size of the region of possible sprite inception to variations in the value of tilt, holding all other parameters constant at their values for our default winter storm. A simulation with zero tilt is also shown for reference. (See Table 2, simulation set 4.) The results are shown in Figure 5.

From Figure 5, we see that the region of possible sprite inception (again, the zero contours) for tilt values of 3.7-5.4 km (dark green, blue, and cyan contours, respectively) are almost indiscernible from one another. The centers of these zero contours are all shifted slightly (by approximately the value of the tilt) from the region of possible sprite inception for zero tilt (black contours, centered at $x = 0$). Thus, the tilt serves to shift the center of the region of possible sprite inception, which means that the electric field in the mesosphere is controlled more by the location of the discharging stroke than by the location of the remaining cloud charge. The more extreme value of tilt of 15.0 km (light green contours) produces a region of possible sprite inception that is shifted markedly from the other contours and that extends out to 60 km laterally from the center of the domain. Optical observations of such a lateral offset of various sprite elements with respect to the location of the parent +CG have indeed been previously reported (Wescott et al., 2001, São-Sabbas et al., 2003), and these results provide a possible explanation for those observations.

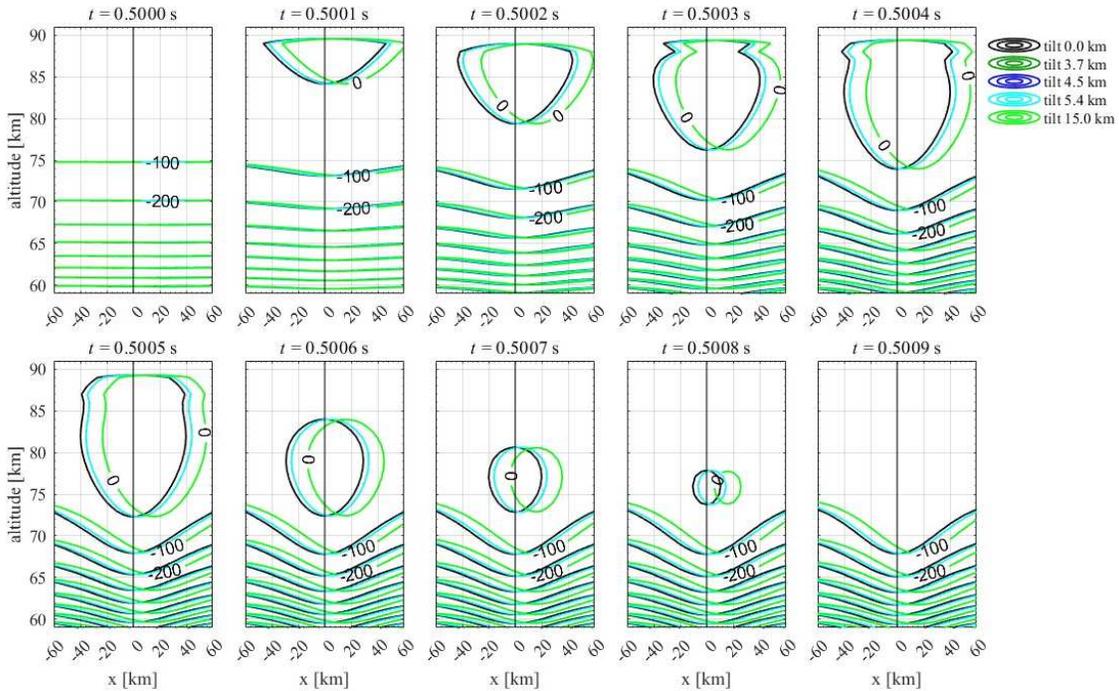

**Figure 5.** Contours of the difference between $\left|\bar{E}(x,y,z,t)\right|$ and $E_{\mathrm{k}}(z)$ in a cross section of the domain at $y = 0$ km for a winter thunderstorm with varying values of $Q_{0_{\max}}$ and with all other parameters held constant at their default values. (See Table 2, simulation set 4).

### 3.5 The effect of lowering the altitude of the base of the ionosphere

The atmosphere in winter is colder than in summer, with different density profiles as well as different temperature profiles (Su et al., 1998), resulting in a lower altitude of the base of the ionosphere in winter. Given that, here we examine the effect of lowering $z_{\mathrm{ionosphere}}$ from 90 km to 87 km and to 85 km, respectively, holding all other parameters constant at their values for our default winter storm. (See Table 2, simulation set 5). The results are shown in Figure 6.

From Figure 6, for lower values of $z_{\mathrm{ionosphere}}$, the region of possible sprite inception is obviously capped at the ionosphere boundary, which decreases the overall region of possible sprite inception (again, the zero contours) despite the identical iCMC values. However, a region of possible sprite inception exists for the same time steps in all three cases. In addition, due to the decrease in the altitude of the maximum of the magnitude of $\bar{E}(x,y,z,t)$ with time following the onset of the lightning discharge, the three regions of possible sprite inception are nearly identical at times $t = 0.5006\text{-}0.5008$ s. Therefore, the altitude of the base of the ionosphere does not influence the region of possible sprite inception 0.0006-0.0008 seconds after the onset of the discharge.

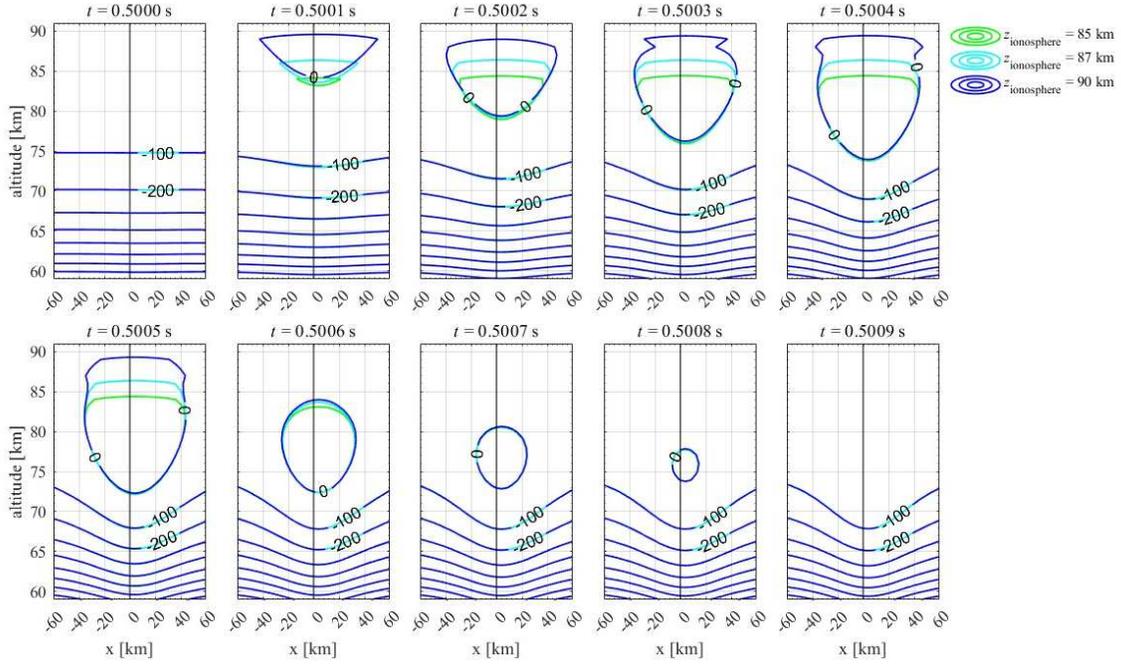

**Figure 6.** Contours of the difference between $\left|\bar{E}(x,y,z,t)\right|$ and $E_{k}(z)$ in a cross section of the domain at $y = 0$ km for a winter thunderstorm with varying values of $z_{\text{ionosphere}}$ and with all other parameters held constant at their default values. (See Table 2, simulation set 5).

### 3.6 The effect of decreasing the altitudes of the cloud charge centers

Though cloud charge center altitudes of 9 and 6 km are representative of eastern Mediterranean winter storm clouds (refer to section 2.2), some winter storm clouds are even more compact. (See, e.g., Yair et al. (2015) and Wang et al. (2018).) To examine the effect of the compactness of a winter storm cloud, we conduct several simulations with cloud charge centers at even lower altitudes (6 km and 3 km, and 8 km and 5 km, respectively), holding all other parameters constant at their values for our default winter storm. (See Table 2, simulation set 6). The results are shown in Figure 7.

From Figure 7, we can see that with the lower cloud charge center altitudes and the corresponding lower iCMC values (green contours and cyan contours, respectively), the regions of possible sprite inception (again, the zero contours) are smaller and dissipate sooner as compared to our default winter storm (blue contours). At $t = 0.5006$ s, there is no region of possible sprite inception with cloud charge center altitudes of 6 km and 3 km, and at $t = 0.5008$ s, there is no region of possible sprite inception with cloud charge center altitudes of 8 km and 5 km.

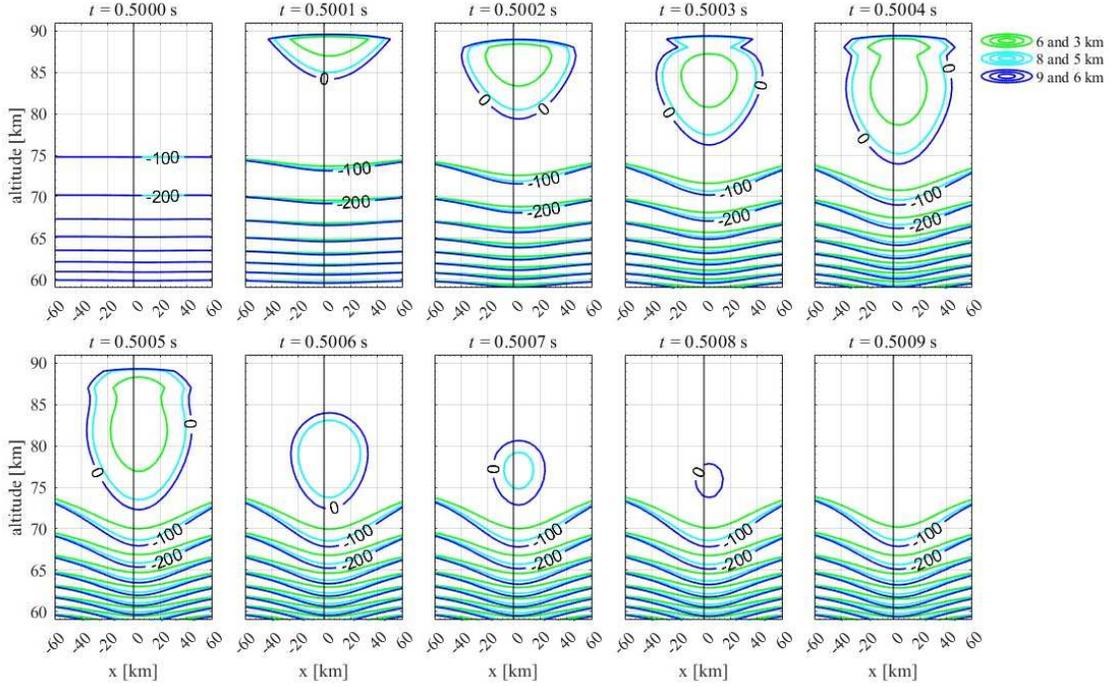

**Figure 7.** Contours of the difference between $|\vec{E}(x,y,z,t)|$ and $E_k(z)$ in a cross section of the domain at $y = 0$ km for a winter thunderstorm with lower values of $z_{\text{upper}}$ and $z_{\text{lower}}$ and with all other parameters held constant at their default values. (See Table 2, simulation set 6).

### 3.7 Consecutive discharges in neighboring winter cloud cells

Although winter thunderstorms generally exhibit low flash rates compared with their summer counterparts, there may be instances when mature cumulonimbus cells will discharge in very close temporal and spatial proximity. This so-called synchronicity was suggested by Yair et al. (2009) based on thunderstorm data in the eastern Mediterranean. Under such circumstances, the quasi-electrostatic field in the mesosphere would be a superposition of the fields generated by each individual discharge. In Haspel et al. (2022), we investigated the possibility of sprite inception due to consecutive discharges from ostensibly summer thunderstorm cloud cells (refer to Table 2, simulation 2) but with $Q_{0_{\max}}$ = 100 C in each cell and an $E_k$ profile corresponding to annual average conditions at mid-latitudes (45°N). The clouds were offset from one another spatially and temporally. We found that the presence of the first cloud cell *in a pre-lightning state* has a minimal influence on the electric field above the second cloud cell, while the presence of the first cloud cell *in a discharging or discharged state* increases the size and duration of the region of possible sprite inception above the second cloud cell.

In the present study, we conduct a similar experiment, but this time on consecutive discharges from neighboring *winter* cloud cells with $Q_{0_{max}}$ = 100 C in each cell. Cloud cell 1 is first to discharge and is offset from the center of the domain and from cloud cell 2 by a distance $D$ in the positive $x$-direction. Cloud cell 2 begins its discharge with a delay time, $t_{delay}$, after cloud cell 1 begins its discharge.

In our first simulation of consecutive discharges from neighboring winter cloud cells, we set $D$ = 20 km and $t_{delay}$ = 1 ms. The simulation parameters for each cloud cell are otherwise identical to those of our default winter cloud cell. (See Table 2, simulation 7). The vertical profiles and vertical cross sections of charge density and electric field in center of the domain for this simulation are shown in Figure 8.

As can be seen in rows 2-4 of the plots of Figure 8, similar to the results shown in Figure 1, there is a clear lack of symmetry in the charge density and in the magnitude of the electric field, and this is due to both the lateral offset of cloud cell 1 from the center of the domain and due to the wind shear experienced by each cloud. As in Figure 1, this lack of symmetry results in peak values of charge density and electric field that are not in the center of the domain. Nevertheless, once again, in row 1 of the plots, we display the vertical profiles of charge density and electric field in the center of the domain for direct comparison/contrast with Pasko et al. (1995, 1997), Haspel et al. (2022), and Figure 1 of the present study. Once again, the lack of symmetry in the charge density and in the magnitude of the electric field is exactly the type of phenomenon that can be captured with our fully 3D simulation rather than with a 2D axisymmetric model.

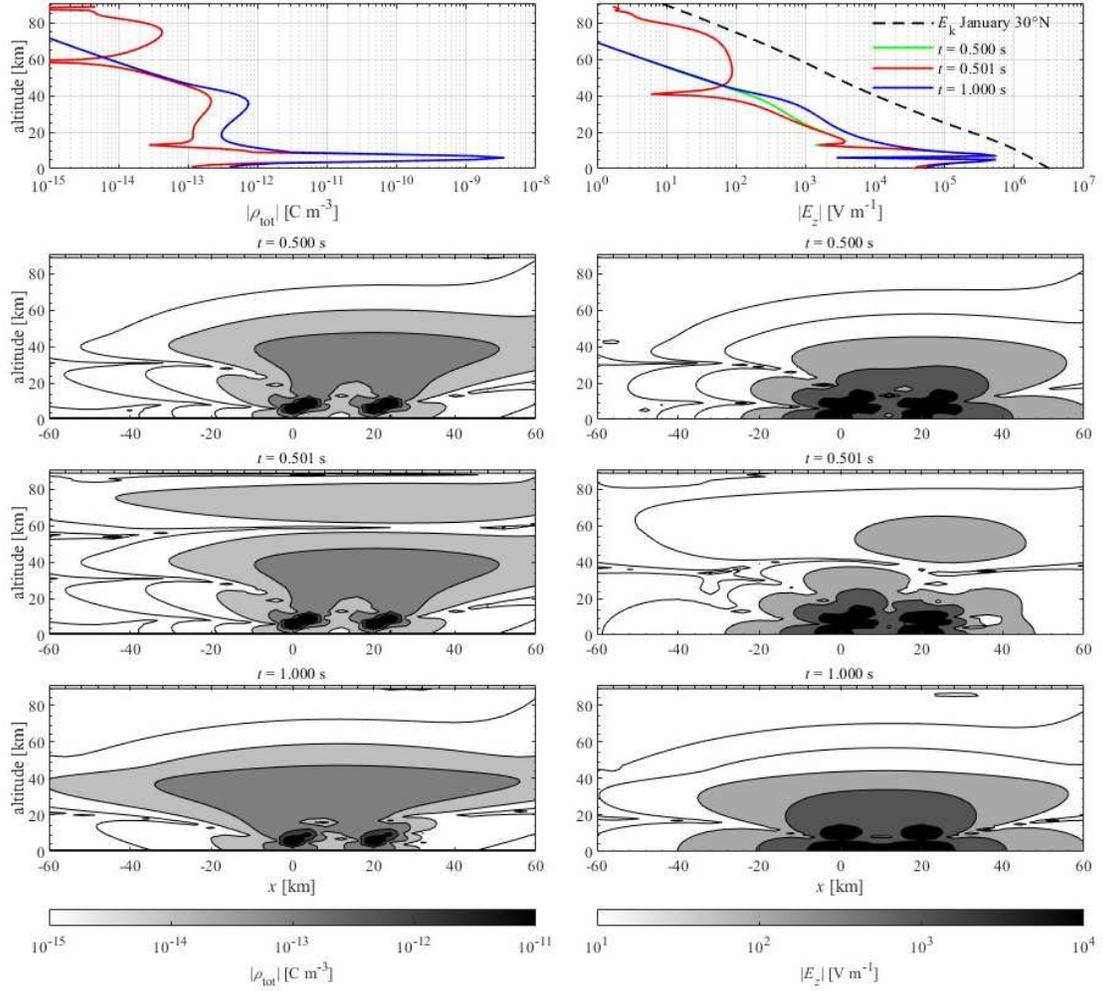

**Figure 8.** Simulation of the vertical profile of total charge density and the $z$-component of the electric field in a cross section of the domain at $y = 0$ km for consecutive discharges in neighboring winter cloud cells with the parameter set given in Table 2, simulation 7.

Contours of the difference between $|\vec{E}(x,y,z,t)|$ and $E_k(z)$ in a cross section of the domain at $y = 0$ km for this simulation with consecutive discharges in neighboring winter cloud cells are shown in Figure 9. From Figure 9, we can see that when cloud cell 1 begins its discharge while cloud cell 2 is still in a pre-lightning state (cyan contours at $t = 0.5001$ s), there is only a tiny region of possible sprite inception (again, the zero contours). This tiny region of possible sprite inception is the same size as for a single winter cloud cell with $Q_{0_{max}}$ = 100 C (black contours; refer also to Figure 4) but shifted 20 km laterally to the location of cloud cell 1. This observation agrees with the first observation from Haspel et al. (2022) stated above, that the presence of the first cloud cell in a *pre-lightning state* has a minimal influence on the electric field above the second cloud cell. Once cloud cell 2 begins its discharge at $t = 0.5010$ s, a larger region of possible sprite inception forms at $t = 0.5011$ s (cyan contours), lasting until $t = 0.5015$ s. Thus, the presence of cloud cell 1 in a *discharging state* influences the electric field above cloud cell 2 during its discharge, increasing both the size and duration of the region of the possible sprite inception above cloud cell 2 and thereby increasing the possibility that a sprite will occur above cloud cell 2. This observation agrees with the second observation from Haspel et al. (2022) stated above, that the presence of the first cloud cell in a *discharging or discharged state* increases the size and duration of the region of possible sprite inception above the second cloud cell. Note that though the total value of $Q_{0_{\max}}$ over the two cloud cells is 200 C, the timing and size of the region of possible sprite inception differs from that for a single cloud cell of 200 C (shown for reference as the blue contours).

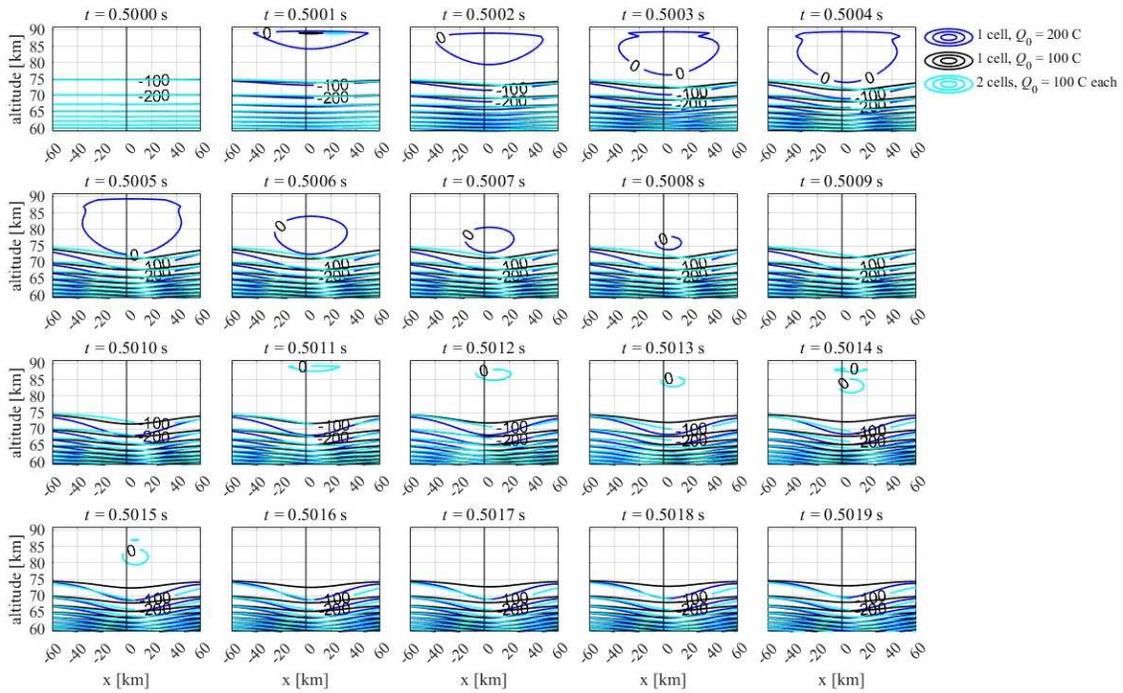

**Figure 9.** Contours of the difference between $|\bar{E}(x,y,z,t)|$ and $E_k(z)$ in a cross section of the domain at $y = 0$ km for consecutive discharges in two neighboring winter cloud cells with $Q_{0_{\max}} = 100$ C in each cell (cyan contours). The cells are offset from one another spatially in the $x$-direction by $D = 20$ km and offset temporally by a delay time of $t_{\text{delay}} = 1$ ms. The simulation parameters for each cloud cell are otherwise identical to those of our default winter cloud cell. (See Table 2, simulation 7.) The result for a single winter cloud cell with $Q_{0_{\max}} = 200$ C (blue contours) and the result for a single winter cloud cell with $Q_{0_{\max}} = 100$ C (black contours) are shown for reference.

We also seek the minimum value of $Q_{0_{max}}$ in each cloud cell at which we still see the influence of one winter cloud cell on the region of possible sprite inception above another winter cloud cell. For consecutive winter storm cells with $D = 20$ km and $t_{delay} = 1$ ms and with all other parameters at their default values, this minimum value of $Q_{0_{max}}$ in each cell turns out to be 90 C. See Table 2, simulation 8 and Figure 10.

From Figure 10, with $Q_{0_{max}} = 90$ C (yellow contours), there is actually *no* region of possible sprite inception when cloud cell 1 begins its discharge at $t = 0.5000$ s, but after cloud cell 2 begins its discharge at $t = 0.5010$ s, there is a tiny region of possible sprite inception at $t = 0.5011$ s centered at $x = 24.5$ km. The result for a single winter cloud cell with $Q_{0_{max}} = 100$ C (black contours) and for two neighboring winter cloud cells with $Q_{0_{max}} = 100$ C from Figure 9 (cyan contours) are also shown for reference. The simulation with $Q_{0_{max}} = 90$ C in each cell provides a possible explanation for a delayed sprite: a sprite might appear after a discharge in a second cloud even when there is no possibility of a sprite above the first cloud, i.e., a delayed sprite can be caused by a "one-two punch".

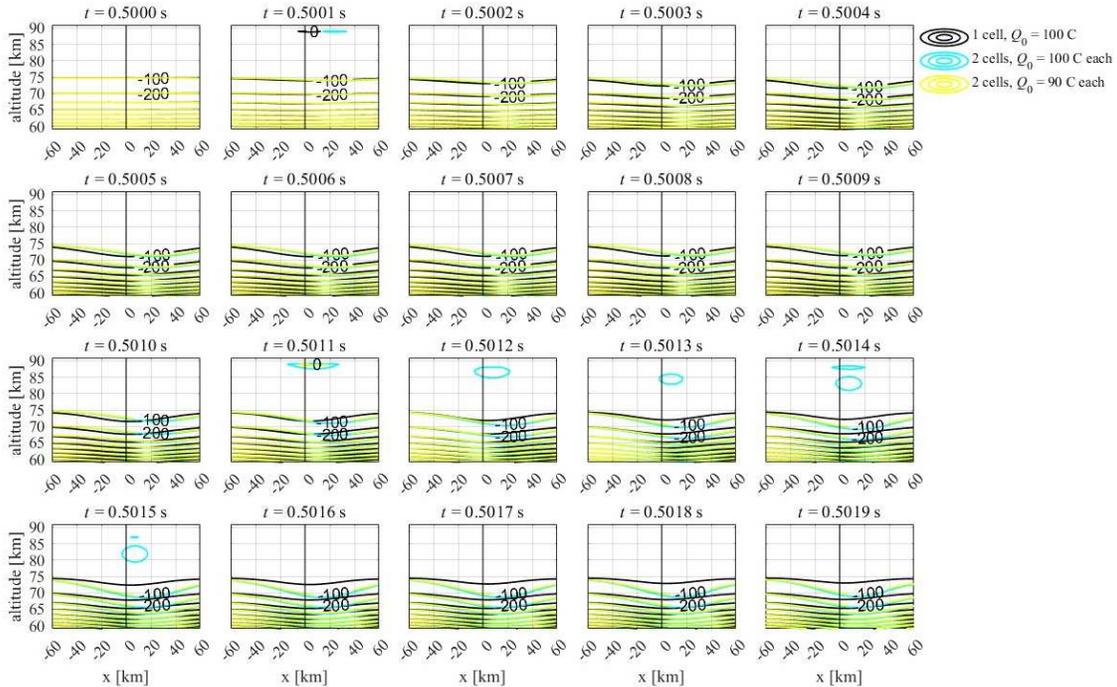

**Figure 10.** Contours of the difference between $|\bar{E}(x,y,z,t)|$ and $E_k(z)$ in a cross section of the domain at $y = 0$ km for consecutive discharges in two neighboring winter cloud cells with $Q_{0_{max}} = 90$ C in each cell (yellow contours). The cells are offset from one another spatially in the $x$-direction by $D = 20$ km and offset temporally by a delay time of $t_{delay} = 1$ ms. The simulation parameters for each cloud cell are otherwise identical to those of our default winter cloud cell. (See Table 2, simulation 8.) The result for a single winter cloud cell with $Q_{0_{max}} = 100$ C (black contours) and for two neighboring winter cloud cells with $Q_{0_{max}} = 100$ C (cyan contours) are shown for reference.

Finally, we seek the maximum value of $D$ at which we still see the influence of one winter cloud cell on the region of possible sprite inception above another winter cloud cell for two neighboring cells with $Q_{0_{max}} = 100$ C. We find that with a delay time of $t_{delay} = 1$ ms, there is still a small influence on the size of the region of possible sprite inception above cloud cell 2 at $t = 0.5011$ s with $D > 200$ km (results not shown here). With $Q_{0_{max}} > 100$ C, the value of $D$ at which we still see the influence of one winter cloud cell on the region of possible sprite inception above another winter cloud cell would be even greater.

## 4. Discussion

One of our overall results from section 3 is that there is a decreased likelihood of sprite inception by a winter thunderstorm as compared to a summer thunderstorm. As mentioned in section 3, this corresponds with regional and global observations (Chern et al., 2003; Yair et al., 2015; Arnone et al., 2020). One major reason for this difference in sprite occurrence above winter and summer storms is the lower CMC and iCMC values resulting from the lower altitudes of the cloud charge centers for the same total charge removed by the parent +CG stroke. From observations of sprites over winter thunderstorms that do occur, the CMC or iCMC tends to be similar or higher than that of summer thunderstorms (Sato and Fukunishi, 2003), which implies that in order to achieve a sprite above a winter thunderstorm, to compensate for the lower altitude of the cloud charge centers, the total charge removed must be higher.

A second possible reason for the difference in sprite occurrence above winter and summer storms is the vertical profile of the conventional electrical breakdown field. The vertical profile of $E_k$ that was adopted for our winter storms exhibits slightly lower values of $E_k$ than that for typical summer conditions at most of the mesospheric altitudes but crosses over the summer profile at altitudes just below 90 km. (Refer to Figure 1 of Haspel et al. (2020)). We found that this factor partially compensates for the lower iCMC.

As described in section 2, for our default winter thunderstorm, we set the radius of the Gaussian "disks" of charge to be 1 km rather than the 3 km used in previous studies to simulate more summer-like thunderstorms (e.g., the simulations in Pasko et al. (1995, 1997)). This is consistent with the smaller dimensions of winter thunderstorms as deduced from observations of storms in Japan and the eastern Mediterranean. However, we found that while this difference affects the electric field near the thunderstorm, it produces no discernable difference in the magnitude of the electric field in the mesosphere. This observation agrees with a result from Haspel et al. (2020) that in the purely static case, we could even use single point charge centers and get a similar result to using Gaussian "disks" of the same total value of charge.

As also described in section 2, we simulated our winter thunderstorms as being comprised of a regular dipole configuration of cloud charge centers. However, it has been suggested that a tripole configuration might be more typical of actual thunderstorms (e.g., Williams, 1989), and some previous studies have simulated such tripole configurations (see, e.g., Riousset et al. (2010)). To test whether adding an additional charge center would affect our results, based on one of the cases investigated in Riousset et al. (2010), we conducted a simulation with a third charge center placed at 2-km altitude with a charge of +10 C and with all other parameters set equal to those of our default winter thunderstorm (refer to Table 2, simulation 1). We found that this third charge center produces some differences in the magnitude of the electric field below 50-km altitude but no discernable difference in the magnitude of the electric field above 50-km altitude and thus no discernable difference in the region of possible sprite inception above the thunderstorm (results not shown here). This result still fits the rule that the value of the CMC or iCMC controls the potential for sprite occurrence, given that the addition of the third charge center does not influence the total charge removed during the discharge. Perhaps just as importantly, the third charge center does not perceptively influence the negative screening charges that accumulate above the upper positive charge center. Other charge configurations, such as an inverted dipole, are beyond the scope of the present study and not simulated here.

As also described in section 2, for simplicity, as in Haspel et al. (2022), in the present study, we choose $\sigma(x,y,z)$ to equal the ion conductivity represented by profile A of Pasko et al. (1997), i.e., $\sigma(x,y,z) = 5.0 \times 10^{-14} \cdot \exp(z/6 \text{ km})$ S m$^{-1}$, independent of time. We also conducted several sensitivity tests with different ambient conductivity profiles. In the first sensitivity test on the conductivity profile, in a similar fashion to Mallios and Pasko (2012), we zeroed out the conductivity within the cloud. As with adding a third cloud charge center mentioned above, zeroing out the conductivity within the cloud produces some differences in the magnitude of the electric field below 50-km altitude but no discernable difference in the magnitude of the electric field above 50-km altitude and thus no discernable difference in the region of possible sprite inception above the thunderstorm (results not shown here). In the second sensitivity test on the conductivity profile, we set the conductivity profile to $\sigma(x,y,z) = 6.0 \times 10^{-13} \cdot \exp(z/11 \text{ km})$ S m$^{-1}$. This profile is based on Holzworth et al. (1985) and ion conductivity profile C of Pasko et al. (1995, 1997) and was also tested in Mallios and Pasko (2012). Such a profile would be considered an extreme conductivity profile at mesospheric altitudes, with conductivity values significantly lower than typical values. On the other hand, very low conductivity values in the mesosphere may temporarily exist during an influx of meteors/meteorites (Zabotin and Wright, 2001) due to the presence of ablation products to which free ions and electron attach, thus decreasing the ambient conductivity. Thus it is interesting to test this hypothetical case. We find that with such a low conductivity in the mesosphere and with all other parameters set equal to those of our default winter thunderstorm (refer to Table 2, simulation 1), the region of possible sprite inception exhibits much more persistence, lasting until after the end of the discharge. (See Figure 11.) In Figure 11, we see that with such a low conductivity profile in the mesosphere, a region of possible sprite inception continues to persist out to $t = 0.5319$ s (more than 20 ms after the end of the discharge, which was at $t = 0.5100$ s). Lowering the conductivity in the mesosphere increases the relaxation time of the field, and this is a dominant factor in promoting the development/persistence of the region of possible sprite inception. (The importance of the relaxation time of the field is discussed, e.g., in Pasko et al. (1995) and in Mallios and Pasko (2012)).

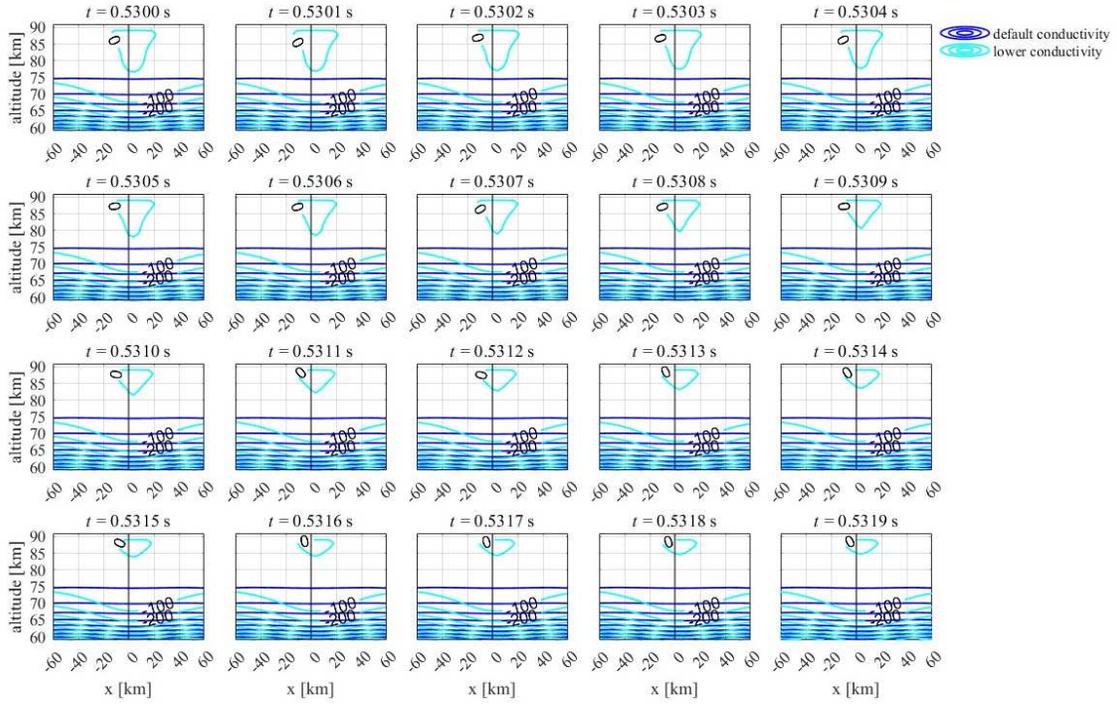

**Figure 11.** Contours of the difference between $|\vec{E}(x,y,z,t)|$ and $E_k(z)$ in a cross section of the domain at $y = 0$ km at times $t = 0.5300$-$0.5319$ s with a lower ambient conductivity in the mesosphere (cyan contours) and with all other parameters held constant at their default values (refer to Table 2, simulation 1). The results with our default ambient conductivity profile are shown for reference as the blue contours.

An interesting and unique aspect revealed in our simulations of consecutive discharges in neighboring winter cloud cells is the remote influence mutually exerted by thunderstorms on the induced electric field in the mesosphere and the potential for sprite occurrence. It is clear that when two (or more) mature cumulonimbus clouds reside in relative proximity, the possibility of near-simultaneous discharges (i.e., "synchronous" discharges; Yair et al., 2006) is non-zero. This may be a good setup for the generation of sprites in areas or times after the discharge for which a single thundercloud would not be sufficient. As in Haspel et al. (2022), we found that the presence of a thundercloud cell in a *discharging state* influences the electric field above a neighboring cloud cell during its own discharge, increasing both the size and duration of the area of the possible sprite inception above it and thereby increasing the possibility that a sprite will occur above a thundercloud that may not be able to produce a sprite on its own. In addition, we found that the largest lateral separation between two winter cloud cells at which we still see the influence of one cell on the area of possible sprite inception of the other cell is > 200 km when the cloud charge magnitude is 100 C. Seemingly, this "one-two" punch to the mesosphere would be enough to induce sprites and can occur regardless of the specific meteorological setting, either in winter or summer storms. In short, when multiple active cells are located within a specific area, the dimensions of which may be of the order of a squall line or mesoscale convective system, the probability for sprite occurrence is enhanced during episodes of flash synchronicity.

**5. Summary and Conclusions**

In this study, we used a 3D quasi-electrostatic model specially designed to handle non-symmetric charge configurations in a large 3D domain (Haspel et al., 2022) to investigate the regions of possible sprite inception above winter thunderstorms exhibiting tilt due to wind shear. To set the configuration of the charges and other simulation parameters, we implemented the characteristics of eastern Mediterranean winter thunderstorms obtained from meteorological observations conducted at specific days when lightning and sprites were optically observed.

We found that the region of possible sprite inception for a winter storm is smaller and exists for shorter periods of time than that for a typical summer storm, which supports the fact that there are less frequent observations of sprites over winter thunderstorms than over summer thunderstorms.

We demonstrated a lack of symmetry in the charge density and in the magnitude of the electric field resulting from the tilt of the winter storm due to wind shear, which is exactly the type of phenomenon that can be captured with our fully 3D simulation rather than with a 2D axisymmetric model. We also found that the tilt of the cloud due to wind shear shifts the center of the region of possible sprite inception by approximately the value of the tilt. For eastern Mediterranean winter thunderstorms, this shift can be up to 15.0 km from the center of the base of the cloud, with the edge of the region of possible sprite inception extending beyond 60 km laterally from the center of the base of the cloud. The role of wind shear may explain the fact that in many cases, sprites appear within 50 km of the ground location of the parent stroke (São-Sabbas et al., 2003).

We found that the minimum value of cloud charge that still produces a region of possible sprite inception with our default winter thunderstorm configuration parameters is ~100 C.

We found that with a lower altitude of the base of the ionosphere, the initial size of the overall region of possible sprite inception is smaller due to being capped by the ionosphere boundary, but the region of possible sprite inception persists for the same duration as when the ionosphere boundary is set at 90-km altitude.

We investigated the effect of decreasing the conductivity in the mesosphere due to the presence of meteoritic dust (Wescott et al., 2001) and found that the region of possible sprite inception persists for longer periods of time, lasting until after the end of the parent lightning discharge, due to the increased relaxation time of the electric field, as discussed, e.g., in Pasko et al. (1995) and in Mallios and Pasko (2012)).

Additionally, we investigated consecutive discharges by neighboring winter cloud cells, in which cloud cell 1 is first to discharge and is offset laterally from the center of the domain and from cloud cell 2, which begins its discharge with a set time delay relative to the discharge from cloud cell 1. We found that as with a single winter cloud cell exhibiting tilt due to wind shear, when there are consecutive discharges in neighboring winter cells, there is a clear lack of symmetry in the charge density and in the magnitude of the electric field, and this is due to both the lateral offset of cloud cell 1 from the center of the domain and due to the wind shear in each cloud. This is again exactly the type of phenomenon that can be captured with our fully 3D simulation rather than with a 2D axisymmetric model.

We found that the minimum value of cloud charge magnitude in each cloud cell at which we still see the influence of one winter cloud cell on the region of possible sprite inception above another winter cloud cell is ~90 C. Moreover, the simulation with a cloud charge magnitude of 90 C in each cell provides a possible explanation for a long-delayed sprite: a sprite might appear after a discharge in a second cloud even when there is no possibility of a sprite above the first cloud, i.e., a delayed sprite can be caused by a "one-two punch".

In these simulations, we adopted a specific discharge time dependence that is a reasonable time dependence considering the overall discharge time we chose of 10 ms. Simulations examining additional discharge time dependence functions will be presented in a separate future study.

Overall, we found that the 3D quasi-electrostatic model of Haspel et al. (2022) is an excellent tool for investigating the qualities of the regions of possible sprite inception above single and multiple winter thunderstorm cells exhibiting tilt due to wind shear.

**Acknowledgements**

Funding: This work was partially funded by the Israel Science Foundation (ISF), Grant 2187/21.

**Tables**

| Shear [m s$^{-1}$ km$^{-1}$] | Tilt [km] |
| --- | --- |
| 4.15 | 3.7 |
| 5.00 | 4.5 (default) |
| 6.00 | 5.4 |
| 16.67 | 15.0 |

**Table 1.** Values of wind shear and corresponding values of tilt (offset of the positive upper cloud charge center in the positive $x$-direction) used in this study.

| Simulation number or simulation set number | 1; default winter thunderstorm | 2 | 3 | 4 | 5 | 6 | 7 | 8 |
|---|---|---|---|---|---|---|---|---|
| $z_{ionosphere}$ [km] | 90 | 90 | 90 | 90 | 85, 87, 90 | 90 | 90 | 90 |
| $x_{max}$ [km] | 60 | 60 | 60 | 60 | 60 | 60 | 60 | 60 |
| $\Delta x$, $\Delta y$ [km] | 4 | 4 | 4 | 4 | 4 | 4 | 4 | 4 |
| $\Delta z$ [km] | 1 | 1 | 1 | 1 | 1 | 1 | 1 | 1 |
| $z_{upper}$ [km] | 9 | 10 | 9 | 9 | 9 | 6, 8, 9 | 9 | 9 |
| $z_{lower}$ [km] | 6 | 5 | 6 | 6 | 6 | 3, 5, 6 | 6 | 6 |
| tilt [km] | 4.5 | 0.0 | 4.5 | 0.0, 3.7, 4.5, 5.4, 15.0 | 4.5 | 4.5 | 4.5 | 4.5 |
| lightning discharge type | +CG | +CG | +CG | +CG | +CG | +CG | +CG | +CG |
| $t_{total\ discharge}$ [ms] | 10 | 1 | 10 | 10 | 10 | 10 | 10 | 10 |
| discharge time dependence | hybrid tanh-linear | pure tanh | hybrid tanh-linear | hybrid tanh-linear | hybrid tanh-linear | hybrid tanh-linear | hybrid tanh-linear | hybrid tanh-linear |
| $\Delta t$ [s] | 0.0001 | 0.0001 | 0.0001 | 0.0001 | 0.0001 | 0.0001 | 0.0001 | 0.0001 |
| $Q_{threshold}$ [C] | 0.000008 | 0.000008 | 0.000008 | 0.000008 | 0.000008 | 0.000008 | 0.000008 | 0.000008 |
| $n_{max}$ | 1 | 1 | 1 | 1 | 1 | 1 | 1 | 1 |
| $Q_{0\,max}$ of each cloud [C] | 200 | 200 | 200, 180, 160, 140, 120, 100 | 200 | 200 | 200 | 100 | 90 |
| iCMC of each cloud [C km] | 1204 | 2000 | 1204, 1084, 963, 843, 722, 602 | 1204 | 1204 | 803, 1070, 1204 | 602 | 542 |
| $E_k$ profile | January | July | January | January | January | January | January | January |
| $D$ [km] | N/A | N/A | N/A | N/A | N/A | N/A | 20 | 20 |
| $t_{delay}$ [ms] | N/A | N/A | N/A | N/A | N/A | N/A | 1 | 1 |

**Table 2.** Summary of the parameters used in the simulations.

**Data Availability**

The computer code used in this research and its output are available from the authors upon request.